\documentstyle[prl,aps,epsfig]{revtex}
\tighten
\draft
\begin{document}
\twocolumn[\hsize\textwidth\columnwidth\hsize\csname 
@twocolumnfalse\endcsname

\title{Critical dynamics of gauge systems: Spontaneous vortex formation in 2D superconductors}
\author{G. J. Stephens$^1$, Lu\'{\i}s M. A. Bettencourt$^2$ and W. H. Zurek$^1$}
\address{$^1$Theoretical Division T-6 MS B288, Los Alamos National Laboratory,
Los Alamos NM 87545}
\address{$^2$Center for Theoretical Physics, Massachusetts Institute of Technology,
Bldg. 6-308, Cambridge MA 02139}

\date{\today}
\maketitle

\begin{abstract}
We examine the formation of vortices during the nonequilibrium relaxation 
of a high-temperature initial state of an Abelian-Higgs system. 
We equilibrate the scalar and gauge fields using gauge-invariant 
Langevin equations and relax the system by instantaneously removing 
thermal fluctuations.  For couplings near critical, 
$\kappa_c=\sqrt{\lambda}/e=1$, 
we observe the formation of large clusters of like-sign magnetic vortices. 
Their appearance has implications for the dynamics of the phase transition, for 
the distribution of topological defects and for late-time phase ordering kinetics.  
We offer explanations for both the observed vortex densities and 
vortex configurations.   
\end{abstract}

\pacs{PACS Numbers : 74.40.+k, 05.70.Fh, 11.27.+d, 98.80.Cq \hfill MIT-CTP-3109 LAUR-01-2579}

\vskip2pc]

Recently much progress has been made in the study of the
dynamics of phase transitions.  The emerging understanding may prove
relevant for a variety of experiments, ranging from the collisions of heavy nuclei at RHIC 
\cite{RHIC} to the sudden cooling of condensed matter systems 
\cite{He3,He4new,Polturak,Carmi,liqcrys,BECs}.  Nonequilibrium
behavior is particularly important for understanding the population of
topological defects which remains after a transition \cite{K,Z}.     

While most previous research focused on global symmetries, 
it is useful to explore systems with (local) gauge symmetry 
\cite{YZ}. This includes critical phenomena in 
the early universe, heavy-ion collisions and superconductors \cite{LiuFrahm}. 
With gauge symmetries, disorder is more subtle
as local phase gradients can be removed by gauge transformations \cite{Geodesic}. 
Moreover, gauge fields contribute  to the (thermo)dynamics 
of the system, potentially leading to new phenomenology.

In this work we examine how critical dynamics is altered in the presence of gauge fields.
We choose a simple and 
controlled physical setting, describing the behavior of long wavelength fields 
in the quench of 2D superconductors. Disorder in the superconducting state 
is associated with the presence of topological configurations (vortices). 
We analyze the formation and dynamics of these defects and trace their 
origins to thermal electromagnetic fluctuations.  The most important result
of this work is the detection of large clusters of like-sign vortices which
form in the wake of a thermal quench. 
 
Hindmarsh and Rajantie (HR) \cite{HR} recently argued  
that gauge fields alter 
the standard Kibble-Zurek predictions for the distribution of topological defects 
formed in a phase transition.  They also observed that magnetic fluctuations can 
create vortex clusters. We provide evidence strongly correlating the  
density and distribution of defects to thermal magnetic fluctuations.
In distinction to HR, we argue that an important role is played by the 
critical magnetic fields of a superconductor and we exhibit the importance of magnetic 
fluctuations on a variety of large length scales in seeding vortex clusters.   

We investigate the dynamics of a complex scalar 
field minimally coupled to electromagnetism, the Abelian-Higgs model, 
in two dimensions.  This provides, for example,
a phenomenological description of a superconducting film. To accurately compare our results 
with experimental films the effects of a small third dimension need to be 
carefully considered.  Nonetheless, this system
provides a tractable and interesting model, allowing for a detailed examination of the importance of
magnetic fluctuations on critical dynamics and defect formation.    
The Lagrangian density is 
\begin{equation}
{\mathcal{L}}= -{1 \over 4}F_{\mu\nu}F^{\mu\nu} +{1 \over 2} |D_\mu \phi|^2 
-{\lambda \over 8} \left( |\phi|^2 - v^2 \right)^2,
\end{equation}
where $\phi$ is a complex scalar field, 
$F_{\mu\nu}=\partial_\mu A_\nu-\partial_\nu A_\mu$ is the field
strength tensor for electromagnetic gauge potential $A_\mu$ and 
$D_\mu =\partial_\mu - i e A_\mu$.
Of the three parameters in the model ($\lambda$, $e$ and $v$) two may be 
rescaled away through coordinate and field redefinitions 
$\phi\rightarrow \phi v$, $A_\mu \rightarrow A_\mu v$, 
$x \rightarrow v \sqrt{\lambda} x$,  
$t \rightarrow v \sqrt{\lambda} t$ and $\beta= v^{-2} \beta$, with 
$\beta= (k_B T)^{-1}$, 
where $k_B$ is Boltzmann's constant.
The dimensionless parameter $\kappa=\frac{\sqrt{\lambda}} {e}$ 
(the ratio of the London penetration depth to the scalar correlation length) 
controls the relative strength of the 
gauge interactions. With these choices the crossover between type-I ($\kappa < 1$) and 
type-II ($\kappa >1$) superconductivity occurs at critical coupling, ${\kappa}_c=1$. 
When $\kappa \rightarrow \infty $ the global phase invariance 
of the theory is recovered.  In the following we work in temporal gauge, $A_0=0$, 
and therefore obtain Gauss' law, 
$\vec \nabla \cdot \vec E = 2 e ~{\rm Im} \left[ \phi^* \pi \right]$,
as a constraint on the evolution. 

To thermalize the system the scalar and gauge fields are coupled
to a reservoir at a temperature $T$. This is achieved by adding dissipative and
stochastic Langevin sources to the equations of motion. 
These cannot break gauge invariance, 
i.e. the Gauss constraint must be preserved, 
resulting, generally, in multiplicative noise and dissipation terms \cite{Krasnitz}.  
There is more than one gauge invariant form for the equations of motion 
and we choose to couple our system to a reservoir through the
observables $\{|\phi|^2,\vec{E}\}$.  
The details of this choice are irrelevant to the state of canonical 
equilibrium reached by the system at long times.

After equilibrating the system, 
we instantaneously remove the noise and evolve with 
overdamped dynamics \cite{Superconductivity},
\begin{eqnarray}
\label{eq-modelA} 
&& \partial_t \phi_a = -\left[ \nabla^2 - e^2 \vert A \vert^2   
- {1 \over 2} (\vert \phi \vert^2 -1) \right]\phi_a 
+ 2 e \epsilon_{ab} A^i \partial_i \phi_b, \nonumber \\
&& \partial_t A_i = -(\nabla \times B)_i +J_i, \\  
&& J_i=- e^2 \vert \phi \vert^2 A_i  
- e \epsilon_{ab} \phi_a \partial_i \phi_b.  \nonumber
\end{eqnarray}
The system of equations (\ref{eq-modelA}), expressed in terms 
of lattice gauge invariant fields \cite{Lattice}, was solved numerically.
The Gauss constraint was satisfied to about $0.1 \%$. 
We used lattice spacing $dx=0.5$ and time step $dt=0.02$.  

\begin{figure}
\begin{center}
\begin{minipage}[t][5.3cm][t]{5.0cm}
\epsfig{figure=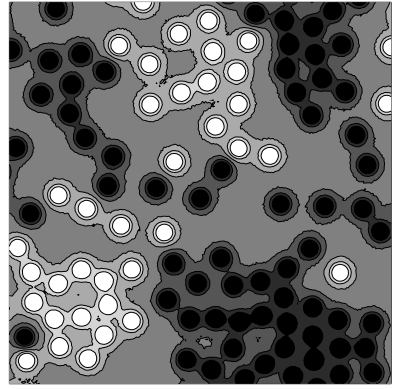,width=2.0in,height=2.0in}
\end{minipage}
\caption{Contour plot of the magnetic flux across a section of the lattice after 
the quench.  White, localized, regions denote vortices while black 
regions denote antivortices. The giant like-sign vortex clusters are   
surprising and are a result of the quench dynamics in a type-II
superconductor, beginning with the formation of 
superconducting regions (in which the gauge fields are small $B<B_{c1}$)
surrounding normal islands of coherent field.
As the quench proceeds the normal islands fragment into clusters of individual
vortices.}
\label{fig:flux}
\end{center}
\end{figure}

A snapshot taken after the quench reveals 
striking structure in the spatial distribution of vortices. 
In Fig.~\ref{fig:flux} we provide a 
contour plot of the magnetic flux across a section of the 
lattice with $e=0.5$ after a quench from   
$T=10$. The critical temperature is $T_c \approx 0.35$.  
The snapshot is late enough in the quench 
that the magnetic flux is 
almost entirely localized into vortices, each of which carries the fundamental 
flux quantum $\Phi_0=\frac{2\pi}{e}$.
The large clusters of like-sign vortices are an obvious departure from
the spatial vortex distribution observed after a quench of  
purely scalar fields.  In the scalar case, defects and antidefects appear 
primarily as widely separated and anticorrelated pairs \cite{Zurek}.  
The appearance of closely-packed vortices is also surprising because 
(though they are not strongly coupled) like-sign vortices repel.

To explain the clustering of vortices we explore the dynamics of the electromagnetic fields.
In equilibrium, at sufficiently high $T$ above the transition,
$\langle B_k B_{-k} \rangle \simeq T$.
When we remove the external noise, all fields dissipate.  
However, not all length scales dissipate at the same rate.  
Long wavelength magnetic fluctuations suffer an analog of 
critical slowing down due to the absence of a magnetic thermal mass in an 
Abelian gauge theory \cite{Fradkin}. 
Under these circumstances we expect
\begin{equation}
\label{eq-Bevolve}
B_k(t) \sim B_k(0) \exp \left[ -{k^2}  t \right].
\end{equation} 
While strictly true only for free dynamics, we verified numerically that 
Eq.~(\ref{eq-Bevolve}) adequately describes the behavior of $B$ in the coupled
system for the initial stages of cooling.  This overdamped evolution results 
in the freezing of magnetic fluctuations on large spatial scales.
Fluctuations in the scalar field are also damped during the cooling.
When fluctuations of the scalar field have decayed sufficiently, 
long-wavelength modes feel the spinodal instability and the scalar field rolls to
its nonzero minimum.  

The evolution of the system 
can be qualitatively understood with reference to the state of minimal 
free energy in an external magnetic field. There are three possibilities
\cite{Superconductivity} delimited by 
\begin{equation}
B_{c1} =  {\Phi_0 \log{\kappa}}/ {4\pi \xi_G^2}, 
\qquad  B_{c2} =  {\Phi_0}/ {2\pi \xi_S^2},
\end{equation}
where $\xi_S$ and $\xi_G$ are the scalar and gauge correlation lengths. 
Regions where $B > B_{c2}$ are forced to remain in the normal phase. 
For $B_{c1} < B < B_{c2}$ the system enters a vortex state, 
where $B$ penetrates a superconducting state in an
array of like-sign vortices. For $B < B_{c1}$ the magnetic field is completely 
expelled and the system becomes a spatially homogeneous superconductor.
In the presence of large thermal $B$, the quench dynamics cause
the system to enter the vortex state in certain regions of space.
Once formed the vortex state is long-lived because
topological charge is conserved and the evolution is overdamped. 
Where $B < B_{c1}$, no vortices are formed by this 
mechanism; the flux is expelled by a Meissner effect and the picture of defect 
formation in terms of the dynamics of the scalar field alone is recovered.
We arrive at a picture of spontaneous vortex formation with two primary qualifications:
\begin{enumerate}
\item The number of vortices formed in the transition is proportional to the amount
of magnetic flux present when the system develops supercurrents.
\item The magnetic field splinters into vortex clusters when $B > B_{c1}$.
The clustered vortex distribution reflects the long-range correlations of the 
initial magnetic field, amplified by the expulsion of flux from regions
where $B<B_{c1}$.
\end{enumerate}
Prediction 1 was first made by HR and our analysis offers quantitative evidence that this 
is correct. However, our second prediction is based on a qualitatively different 
picture of the dynamics of the gauge fields and order parameter during the phase transition.

To test the first prediction we examine the magnetic flux,
$|\Phi|  =\int \sqrt{B^2} d^2x$.
If we replace $B^2$ by its average value then
\begin{equation}
|\Phi| \sim \sqrt{<\!B^2\!>} A,
\end{equation}
where A is the total area of the film.
The determination of the time when supercurrents develop and therefore the point
at which to evaluate $\sqrt{<\!B^2\!>}$ depends on the coupled dynamics of the scalar and
electromagnetic field and will be addressed elsewhere.  However, we can make the following 
(parametric) prediction \cite{HR},
\begin{equation}
\label{eq-estimate}
\rho_{\rm def} = C e \sqrt{T},
\end{equation}
where $\rho_{\rm def}$ is the (area) density of vortices and $T$ is the initial temperature. 
$C$ carries dimensions of inverse length and parameterizes the details  
of the cooling dynamics. In practice we found $C \sim 0.01$ demonstrating that 
only a small fraction of the original thermal magnetic energy is available 
for vortex formation.  We expect Eq.~(\ref{eq-estimate}) to hold only when
the magnetic field at the time of formation is greater than $B_{c1}$.

\begin{figure}
\begin{minipage}[t][7.5cm][t]{5cm}
\epsfig{figure=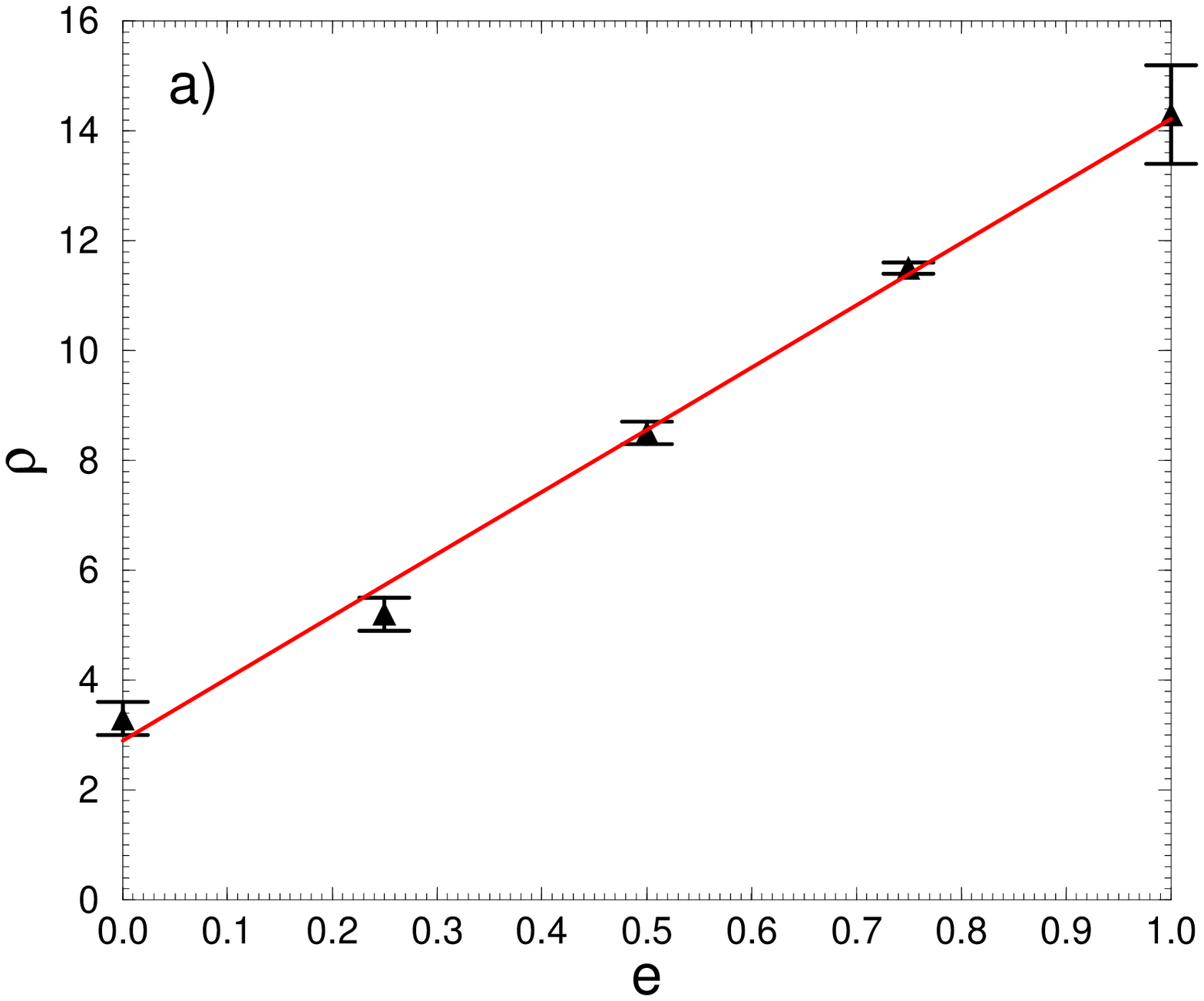,width=7.0cm,height=4.0cm}
\begin{minipage}[t][3.1cm][b]{5cm}
\epsfig{figure=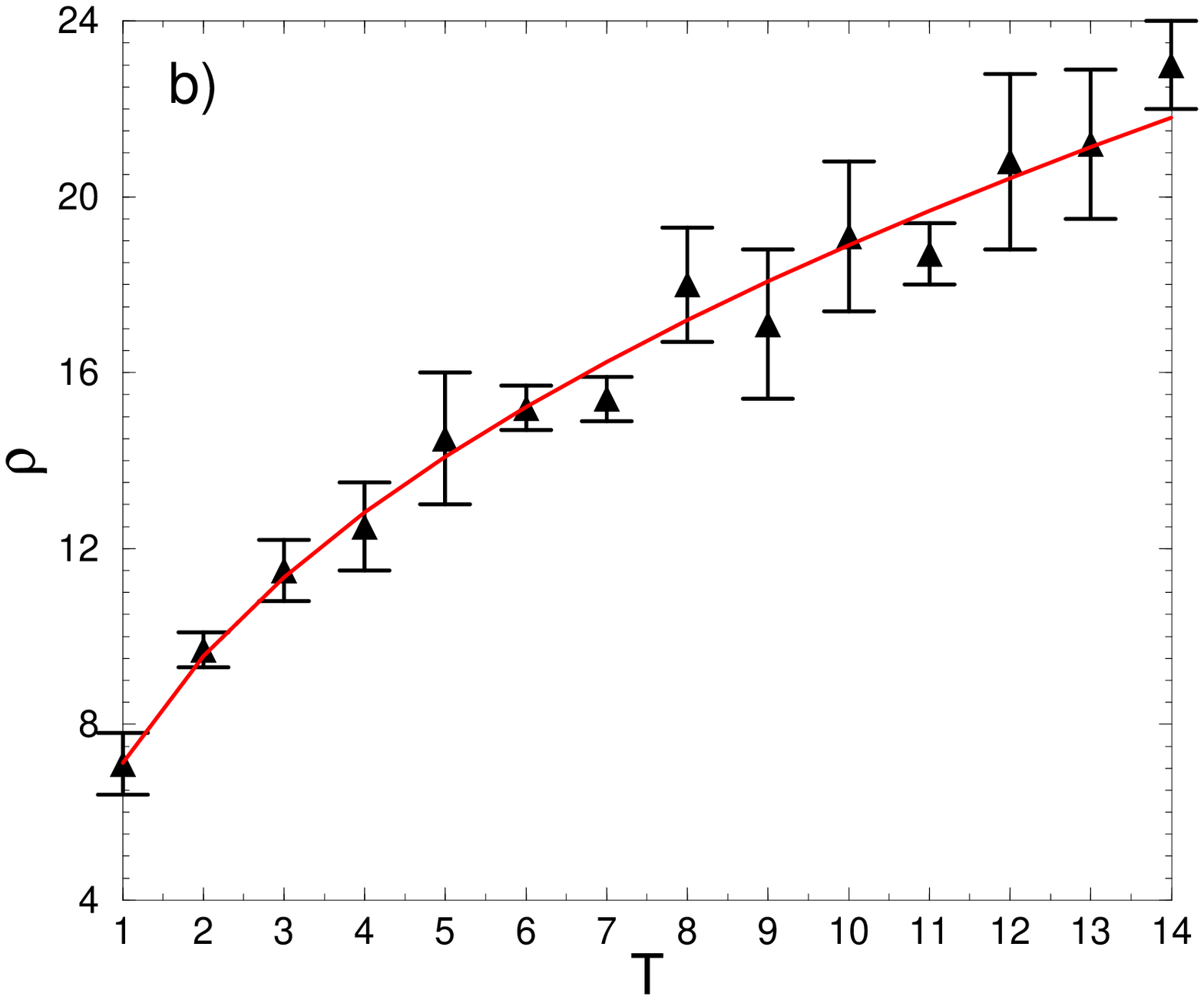,width=7.0cm,height=4.0cm} 
\end{minipage}
\end{minipage}
\caption{a) $\rho_{\rm def} \times 10^{3}$ 
{\it vs.} $e$ from numerical evolutions. The solid line is the best 
linear fit. b) $\rho_{\rm def} \times 10^{3}$ {\it vs.} 
$T$ from numerical simulations.
The solid line is the best power-law fit $\rho \sim T^\alpha$, where
$\alpha=0.43 \pm 0.02$.  In both plots error bars denote standard deviations
among stochastic realizations.}
\label{fig:eTplot}
\end{figure}

\noindent Figs.~\ref{fig:eTplot}a and \ref{fig:eTplot}b show $\rho_{\rm def}$ {\it vs.}  
$e$ and the initial temperature $T$ respectively. Both are in agreement 
with Eq.~(\ref{eq-estimate}).  The slightly smaller exponent in Fig.~\ref{fig:eTplot}b is
an indication that some fluctuations decay before vortices are formed. 

\begin{figure}
\begin{minipage}[t][7.5cm][t]{5cm} 
\epsfig{figure=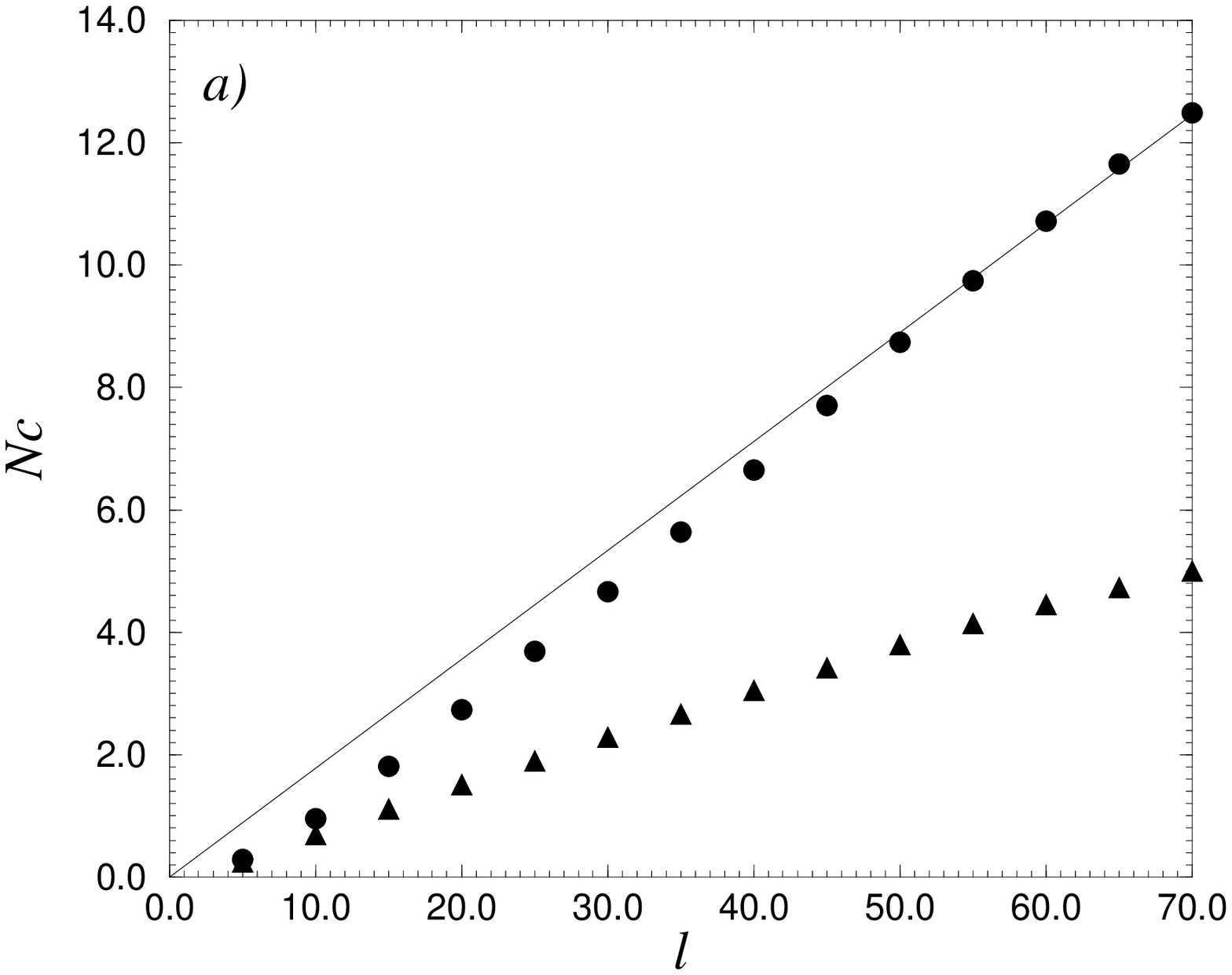,width=7.0cm,height=4.0cm}
\begin{minipage}[t][3.3cm][b]{5cm} 
\epsfig{figure=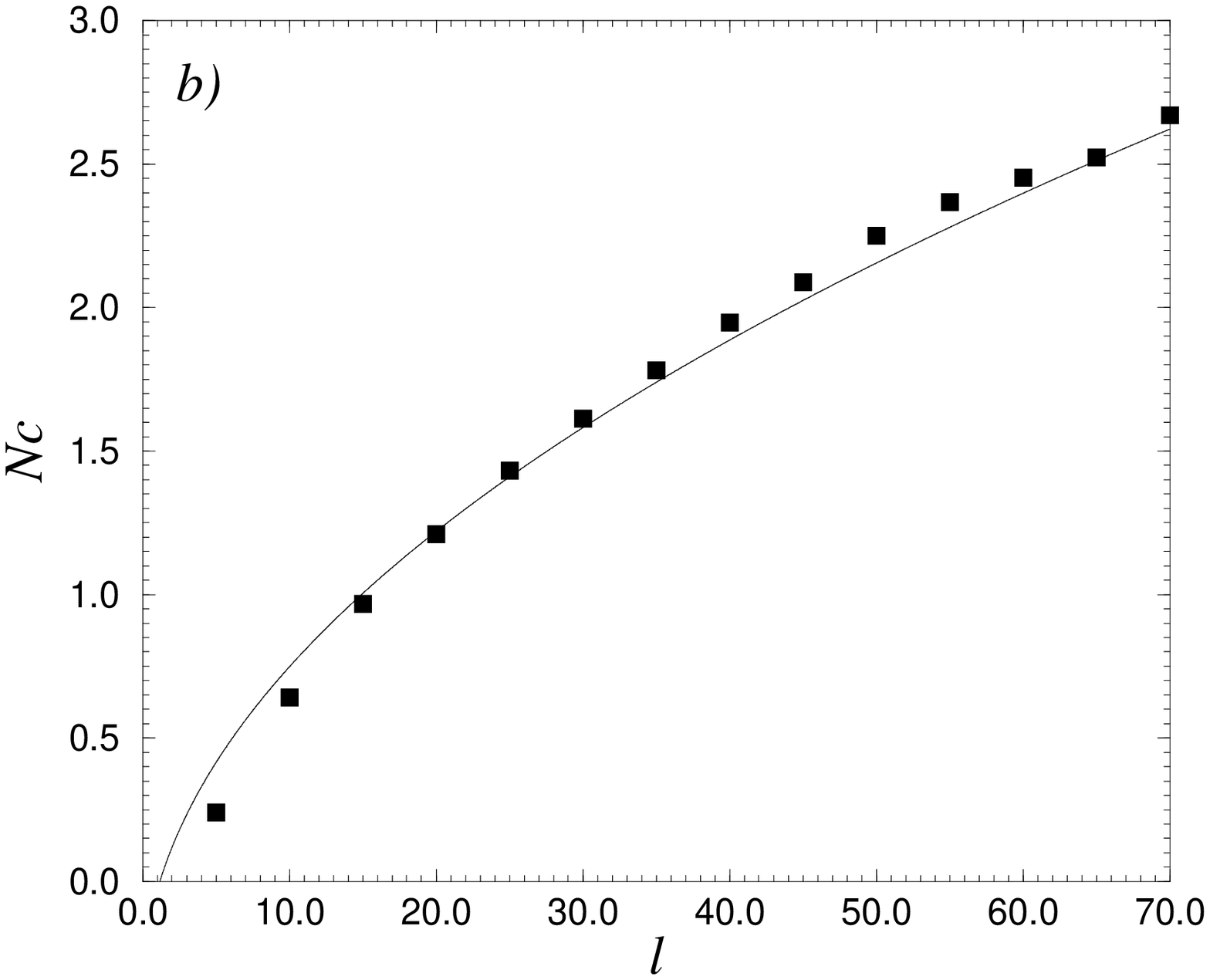,width=7.0cm,height=4.0cm} 
\end{minipage}
\end{minipage} 
\caption{a) $N_c(l)$ for the cluster distribution of 
Fig.~\ref{fig:flux} (circles) compared against a random distribution with 
the same number of defects (triangles). The solid line shows 
the behavior expected from equilibrium fluctuations in the magnetic flux, 
$\Phi_{net}=\sqrt{T} l$.  The close agreement between $N_c(l)$ measured from
the vortex distribution and $N_c(l)$ calculated from equilibrium is strong evidence that
vortex clusters in the gauge quench originate in equilibrium magnetic fluctuations.  
b) $N_c(l)$ for the purely scalar theory, $e=0$, 
normalized to the same density as Fig.~3a.  The solid line is the best 
fit to $\sqrt{l}$} 
\label{fig:netphi1}
\end{figure}

While the {\it number} of vortices depends on the details of the cooling dynamics, 
the clustered {\it distribution} of vortices 
reflects initial magnetic thermal fluctuations on many scales.  During the quench,
magnetic fields are first damped and then redistributed by the Meissner effect: expelled
where $B<B_{c1}$ into initially normal magnetized islands, which eventually fragment into vortices.   
To quantify the properties of the vortex 
cluster distribution we examine the average excess topological charge $N_c(l)$ 
within a square of side $l$.  $N_c(l)$ can be measured in experiments 
using e.g. a SQUID, sensitive to flux over an area $l^2$. 
To calculate $N_c(l)$ we compute the absolute value of the net topological charge in a box of side 
length $l$ centered on a lattice point.  We then average over every lattice point. Fig.~\ref{fig:netphi1} 
shows the result for a typical run with $e=0.5$.  
$N_c(l)$ observed for the gauge quench is indicative of 
clustering and is larger than that of a 
random distribution of the same number of defects.
At small $l$ the presence of clusters of like-sign vortices  
leads to faster than linear growth of $N_c(l)$.
At large $l$ both distributions increase linearly.  This is expected when defects are positioned 
at random: with average density $\rho_{\rm def}$ fluctuations are of order $\sqrt{N(l)}$ 
where $N(l)$ is the average number of defects in box of size $l$.  Therefore
\begin{equation}
N_c^{\rm random}(l)  =\sqrt{\rho_{\rm def}}\: l.
\end{equation}
The linear behavior and larger slope of $N_c(l)$ measured from a gauge quench suggest
that {\it clusters} of like-sign defects are randomly distributed.

In scalar systems defect distributions are not  
random \cite{Zurek}.  Fluctuations of the net topological number over an area occur due to the 
fluctuations of the field phase along the boundary.  After a quench, the phase is 
correlated on a scale $\xi$ with 
$l/\xi$ separate domains along a path of length $l$. Thus, 
\begin{equation}
\label{eq:scalarnc}
N_c(l) \sim (\rho_{\rm def})^{1/4} \sqrt{l}.
\end{equation}
Fig.~\ref{fig:netphi1}b shows $N_c(l)$ measured from simulations of 
a purely scalar theory. Eq.~(\ref{eq:scalarnc}) has also been verified experimentally in
nematic liquid crystals \cite{Nematic} and is 
roughly consistent with measured fluctuations in
the net flux trapped within a loop of Josephson junctions \cite{Carmi}.  

\begin{figure}
\begin{center}
\epsfig{figure=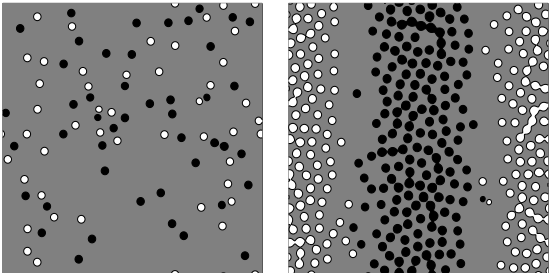,width=3.0in,height=1.5in}
\caption{Contour plot of the magnetic flux after 
a quench in an external magnetic field.  The left plot corresponds to
amplitude $B_0=0.01$.  The right plot is for amplitude $B_0=0.16$.  The 
lower critical field is $B_{c1}=0.08$ while $B_{c2}=0.5$.}  
\label{fig:extB}
\end{center}
\end{figure}

It is instructive to examine defect formation in the presence of
an {\it imposed} magnetic field (see also \cite{Rajantie}).  We prepare an uncoupled 
initial state in which the scalar field is thermalized to a temperature close
to $T_c$ and the magnetic field is 
\begin{equation}
B(t=0)=B_0 \cos \left (\frac{2\pi x}{L} \right ).
\end{equation}
At the time of the quench we turn on the coupling (to $e=0.5$) and evolve
with the overdamped equations (\ref{eq-modelA}).  
The results for two different $B_0$ are shown in 
Fig.~\ref{fig:extB}.  
\noindent For small amplitudes, the final defect distribution 
is not heavily influenced by the magnetic field and large vortex clusters 
are absent. In contrast for $B > B_{c1}$ the magnetic field has a profound 
influence on the final defect distribution. 
These simple considerations mimic the behavior 
occurring on {\it many} length scales in the thermal quench.

In conclusion, we have shown that the critical dynamics of 2D superconductors 
differs substantially from scalar field models.
Specifically we observed that for quenches from high $T$ and for $\kappa \stackrel{>}{{}_\sim} \kappa_c$, 
remnants of long-wavelength magnetic fields seed the formation of topological 
defects in much larger numbers and in strikingly different configurations than 
scalar field theories.  
We established the dependence of the density of vortices on 
$e$ and the initial temperature $T$ and we    
connected the large-scale properties of the vortex distribution
to the spectrum of initial thermal magnetic fluctuations. 
We also argued for the importance of $B_{c1}$  
in determining the effect of magnetic fluctuations upon defect formation. 
With recent experimental progress in vortex imaging \cite{Imaging}, 
clusters such as Fig.~\ref{fig:flux} might be discernible in 
superconductors with $\kappa \sim \kappa_c$ (e.~g.~Nb). 
A detailed analysis of the statistical properties of these clusters 
is under investigation.

We thank M. Hindmarsh, A. Rajantie and N. Goldenfeld for useful
comments.  This work was partly supported by D.O.E grant DF-FC02-94ER40818.  
Numerical work was performed on the T/CNLS Avalon
cluster at LANL.


\begin{thebibliography}{99}




\bibitem{RHIC} See e.g. B.B. Back {\it et al}. Phys.Rev.Lett. {\bf 85} 3100 (2000).

\bibitem{He3} C. Ba\"uerle {\it et al.}, Nature {\bf  382}, 332 (1996);
V.M.H. Ruutu {\it et al.}, Nature {\bf  382}, 334  (1996); 
V.M.H. Ruutu {\it et al.}, Phys. Rev. Lett. {\bf 80}, 1465(1998).

\bibitem{He4new}  M.~E. Dodd {\it et al.},
Phys. Rev. Lett., {\bf 81}, 3703 (1998).

\bibitem{Polturak} R. Carmi, and E. Polturak, Phys. Rev. B 
{\bf 60} 7595 (1999). 
\bibitem{Carmi} R. Carmi, E. Polturak, and G. Koren, 
Phys. Rev. Lett. {\bf 84}, 4966 (2000).

\bibitem{liqcrys} I. Chuang, R. Durrer, N. Turok, and B. Yurke, 
Science {\bf 251}, 1336 (1991); M.J. Bowick, L. Chandar, E.A. Schiff, 
and A.M. Srivastava, Science {\bf 263}, 943 (1994). 

\bibitem{BECs} See e.g. K. Huang, cond-mat/0012418;
 J.~R.~Anglin and W.~H.~Zurek, Phys. Rev. Lett. {\bf 83} 1707 (1999).

\bibitem{K} T. W. B. Kibble, J. Phys.: Math. Gen. {\bf 9} 1387 (1976).
\bibitem{Z} W. H. Zurek, Nature, {\bf 317} 505 (1985).


\bibitem{YZ} A.~Yates, and W.~H.~Zurek, Phys. Rev. Lett. {\bf 80}, 5477 (1998); 
D.~Ibaceta, and E.~Calzetta, Phys. Rev. E {\bf 60} 2999 (1999). In
these works, magnetic fluctuations were too small to form vortex clusters. 

\bibitem{LiuFrahm} F. Liu, M. Mondello and N. Goldenfeld, Phys. Rev. Lett. {\bf 66} 3071 (1991);
H. Frahm, S. Ullah and A. T. Dorsey, Phys. Rev. Lett. {\bf 66} 3067 (1991).

\bibitem{Geodesic} S. Rudaz and A. M. Srivastava, Mod. Phys. Lett. {\bf A8} 1443 (1993);
L. Pogosian and T. Vachaspati, Phys. Lett. {\bf B423} 45 (1998).

\bibitem{HR} M.~Hindmarsh, and A.~Rajantie, 
Phys. Rev. Lett. {\bf 85} 4660 (2000).

\bibitem{Krasnitz} A. Krasnitz, Nucl. Phys. B {\bf 455} 320 (1995).

\bibitem{Lattice} See e.g., M.~Creutz, {\it Quarks, gluons and lattices} 
(Cambridge University Press, Cambridge, U.K., 1985).   

\bibitem{Fradkin} E. Fradkin, Proc.~Lebedev~Phys.~Inst. {\bf 29} 7 (1965).

\bibitem{Superconductivity} See e.g., M.~Tinkham, {\it Introduction to Superconductivity, 
2nd. Edition} 
(McGraw-Hill Higher Education, New York, U.S.A., 1995).   

\bibitem{Zurek}  W.~H. Zurek, Phys. Rep. {\bf  276} 177 (1996).

\bibitem{Nematic} S. Digal, R. Ray and A. M. Srivastava, Phys. Rev. Lett. {\bf 83} 5030 (1999).

\bibitem{Rajantie}  A. Rajantie, J. Low Temp. Phys. {\bf 124} 5 (2001).

\bibitem{Imaging} See e.g. S. H. Pan {\it et al.}, cond-mat/0005484.  

\end{thebibliography}
\end{document}